# Net drag reduction in turbulent pipe flow by airfoil-section grids concentrated near the wall

Takuya Masuda (益田 卓哉)

Department of Integrated Engineering, National Institute of Technology, Yonago College

Hikona-cho 4448, Yonago, Tottori 683-8502, Japan

masuda@yonago-k.ac.jp

## ABSTRACT

The frictional drag of turbulent pipe flow is responsible for enormous energy consumption in water and natural gas distribution systems. Flow straighteners can reduce this drag by suppressing turbulence, but they generally induce a pressure loss that exceeds the reduction; similarly, large-eddy breakup devices installed in pipes have not achieved a net drag reduction. In this study, a grid of NACA0006 airfoils was adopted as a flow straightener to suppress the pressure loss, and its net drag reduction was evaluated by numerical simulations using the $k$–$\omega$ shear stress transport model of the axisymmetric Reynolds-averaged Navier–Stokes equations with periodic boundary conditions. The model was validated against fully developed turbulent pipe flow and a previous experiment on a large-eddy breakup device in a pipe. Under a pressure difference corresponding to a Reynolds number ($Re$) of approximately $10^5$, a grid pattern of concentric rings concentrated near the wall at an axial spacing of 0.5 m increased the flow rate by up to 9.0%, and the flow rate increase remained positive for $5 \times 10^4 \leq Re \leq 2 \times 10^5$. The grid pattern reduced the area-integrated Reynolds shear stress by 29% and brought the velocity profile closer to a laminar one. The effectiveness of the near-wall placement is consistent with the Fukagata–Iwamoto–Kasagi identity, which weights the Reynolds shear stress more heavily closer to the wall.

# I. INTRODUCTION

Pipelines are fundamental infrastructure for water distribution and natural gas supply systems. The transport of a fluid through a pipeline produces frictional drag along its walls, which leads to energy losses. Such energy losses are particularly pronounced in large-diameter pipelines transporting fluid over long distances, where drag is estimated to account for 90% or more of the total energy consumption.[1] Much of the power consumption by pumps and compressors is spent overcoming the wall shear stress, which raises operating costs and increases the greenhouse gas emissions associated with energy production. For example, water distribution is reportedly responsible for approximately 7% of global electricity consumption.[2] Meanwhile, the power loss in the Portuguese high-pressure natural gas transmission network has been estimated at approximately 4%–5% of the transmitted power, with the pressure drop due to viscous friction being a major contributor.[3] Therefore, even a slight reduction in drag would have substantial economic benefits and improve sustainability.

Various approaches have been proposed to reduce the drag on a pipe flow, which can be broadly classified as passive, active, or additive-based.[4,5] Passive methods include surface modifications such as riblets inspired by shark skin, hydrophobic or low-friction coatings, and optimization of the pipe geometry.[6] Active methods such as wall blowing and suction, electromagnetic forcing, and wall oscillation have shown promise in laboratory settings, but their large-scale deployment in industrial piping systems remains challenging because of their complexity and high energy cost.[7,8] In contrast, polymer additives that exploit the Toms effect to reduce drag are already being used in oil pipelines.[9,10] However, further research is needed on robust and economically viable drag reduction technologies to address challenges regarding durability, scalability, and cost-effectiveness.[11]

Drag depends strongly on the flow state as a turbulent flow produces a markedly larger wall shear stress than laminar flow. In a turbulent flow, three-dimensional vortical structures enhance the wall-normal transport of momentum and increase the mean velocity gradient, which increases the drag. Conversely, suppressing these turbulent structures to bring the flow closer to a laminar state can substantially lower the drag. Riblets have attracted considerable attention because they suppress the spanwise motion of near-wall vortices and can reduce drag by up to 10% under optimal conditions.[6]

Large-eddy breakup (LEBU) devices are passive manipulators consisting of a thin plate or airfoil placed near the wall along the flow direction so that its wake breaks up and weakens large-scale vortical structures.[12,13] Research has been mainly experimental although numerical approaches have also appeared in recent years. Spalart et al.[14] and Kim et al.[15] performed direct numerical simulations (DNSs) while Chin et al.[16,17] and Chan et al.[18,19] performed large-eddy simulations (LESs) on LEBU devices. However, all of them were limited to Reynolds numbers lower than practical levels. Meanwhile, to the best of the author's knowledge, Reynolds-averaged Navier–Stokes (RANS) simulations of LEBU devices have been limited to a single study by Shirose[20] more than 30 years ago.

Most studies on LEBU devices have targeted the turbulent boundary layer over a flat plate.[21–23] Although some studies have placed LEBU devices in pipes, the mainstream conclusion has been that they reduce the local wall shear stress but yield no net drag reduction once the form drag of the LEBU device itself is included, and the pressure loss increases. Pollard et al.[24] performed many experiments and calculations on applying LEBU devices of various shapes to turbulent pipe flows. They concluded that a net drag reduction may be possible but would be difficult at present. Mah et al.[25] installed single and tandem blades in a fully developed pipe flow. Based on the changes to the volumetric flow rate and wall pressure distribution far downstream, they found that the single blade increased the net drag, and that the tandem blades were even worse. They concluded that blades have low applicability to pipe flows. Khoo et al.[26] analyzed the skin friction distributions for the same configurations as Mah et al.[25] and showed that both the single and tandem blades increased the total drag of the pipe flow. They concluded that a single LEBU device has little potential for pipe applications. A review by Alfredsson and Örlü[13] found that LEBU devices can decrease the local skin friction coefficient in external boundary layers but generally do not result in a net drag reduction, in either external or internal flows.

Flow straighteners such as honeycombs and wire meshes have long been used to rectify velocity nonuniformity and vortices at the pipe inlet, and they may also rectify nonuniformity in the mean shear stress. Yin et al.[27] conducted a numerical study on a double-bend pipe and reported that installing a flow straightener to reduce turbulence at the inlet resulted in a mild improvement in the downstream pressure loss. However, the flow straightener itself induces a pressure loss, so the key to the design is whether the gain from its rectifying effect exceeds this loss. Moreover, the rectifying effect may be limited in the turbulent regime as the effectiveness of a flow straightener depends on the Reynolds number, inlet turbulence intensity, and pipe geometry (e.g., bends, changes in cross-section). Jurga et al.[28] varied the thickness ratio and installation distance of a flow straightener downstream of a bend and examined the tradeoff between the flow recovery distance and drag reduction. Kühnen et al.[29] used 3D-printed honeycombs to manipulate the inlet velocity profile of a straight circular pipe and relaminarize a turbulent flow. They compared the pressure loss caused by adding the flow straightener with the reduction in the downstream pressure loss. Their results showed that a large reduction in the downstream pressure loss could be achieved.

However, complete relaminarization was achieved only up to a Reynolds number of about 10,000. In addition, their method requires the pipe downstream of the flow straightener to be straight and smooth.

Given this background, a flow straightener should be able to reduce the net drag on a pipe flow if the pressure loss due to its form drag can be suppressed. This study investigated the effect of using a grid of thin airfoils as a flow straightener on the net drag of a pipe flow. Numerical simulations were conducted to optimize the shape and arrangement of the grid and to elucidate the operating principle.

## II. METHODS

### A. Strategy

The NACA0006 airfoil was adopted for the cross-section of the flow straightener in the streamwise direction. It has a thickness that is only 6% of the chord length. Conventional flow straighteners have circular or rectangular cross-sections in the streamwise direction. Even with the same thickness, their form drag is several times that of the NACA0006 airfoil, and they may even disturb the flow through Kármán vortex shedding, which makes them doubly unsuitable for drag reduction. A two-dimensional analysis can be applied by arranging the airfoils in a grid pattern of concentric rings. A practical product would combine concentric rings with radial spokes. The angle of attack was set to 0° to prevent deformation due to lift. Grids were arranged periodically along the streamwise direction. Under these conditions, the optimal values for the chord length $c$, number of rings $n$, radial ring spacing $s$, and axial grid spacing (pipe length) $L$ were investigated.

The Fukagata–Iwamoto–Kasagi (FIK) identity[30] is an exact relation that connects the wall shear stress, which is expressed in terms of the skin friction coefficient $C_f$, to the Reynolds shear stress distribution, which for a fully developed pipe flow reads:

$$C_f = \frac{16}{Re_b} + 16\int_0^1 2r^*\overline{u_r^{*\prime}u_z^{*\prime}}r^*dr^*, \tag{1}$$

where the asterisk (*) denotes a dimensionless quantity. Quantities are nondimensionalized by the pipe radius $D/2$ and twice the mean velocity $U_m$ with $r^* = 0$ corresponding to the pipe center and $r^* = 1$ corresponding to the pipe wall. The bulk Reynolds number $Re_b$ is identical to the Reynolds number $Re = U_mD/\nu$, which is defined without the grid. Two points follow from Eq. (1): (i) suppressing turbulent fluctuations is effective for reducing drag, and (ii) the closer the suppression is to the pipe wall, the greater its effect. The latter is particularly important in designing the grid pattern. However, Eq. (1) assumes a fully developed flow, so it cannot be applied directly to the region downstream of the grid.

B. Target application

The flow straightener was assumed to be applied to a water supply system. The distribution mains of a water supply system buried underneath a road network typically have nominal diameters of 100, 200, or 400 mm and a mean flow velocity of 0.5–2 m/s, which result in $Re$ values of $10^5$–$10^6$. Therefore, $D = 100$ mm, $U_m \approx 1$ m/s, and a kinematic viscosity of $\nu = 10^{-6}$ m$^2$/s were taken as representative, which correspond to $Re \approx 10^5$. Incidentally, buried gas pipes have diameters of 100–300 mm and mean velocities of 5–10 m/s, but the kinematic viscosity is also an order of magnitude larger, so $Re$ is again $10^5$–$10^6$. Small-diameter service pipes connecting to buildings can also be targeted, but they already have a small drag loss and long periods of no flow, so the drag reduction of the flow straightener would probably not justify its installation cost.

C. Numerical model

Figure 1 shows a schematic of the analysis target. The fluid flows in the positive $z$ direction of the cylindrical coordinate system. The flow straightener was modeled as a grid of concentric rings whose center coincides with the origin O. The radius of the innermost ring is denoted by $r_1$, and that of the $i$-th ring from the inside is denoted by $r_i$.

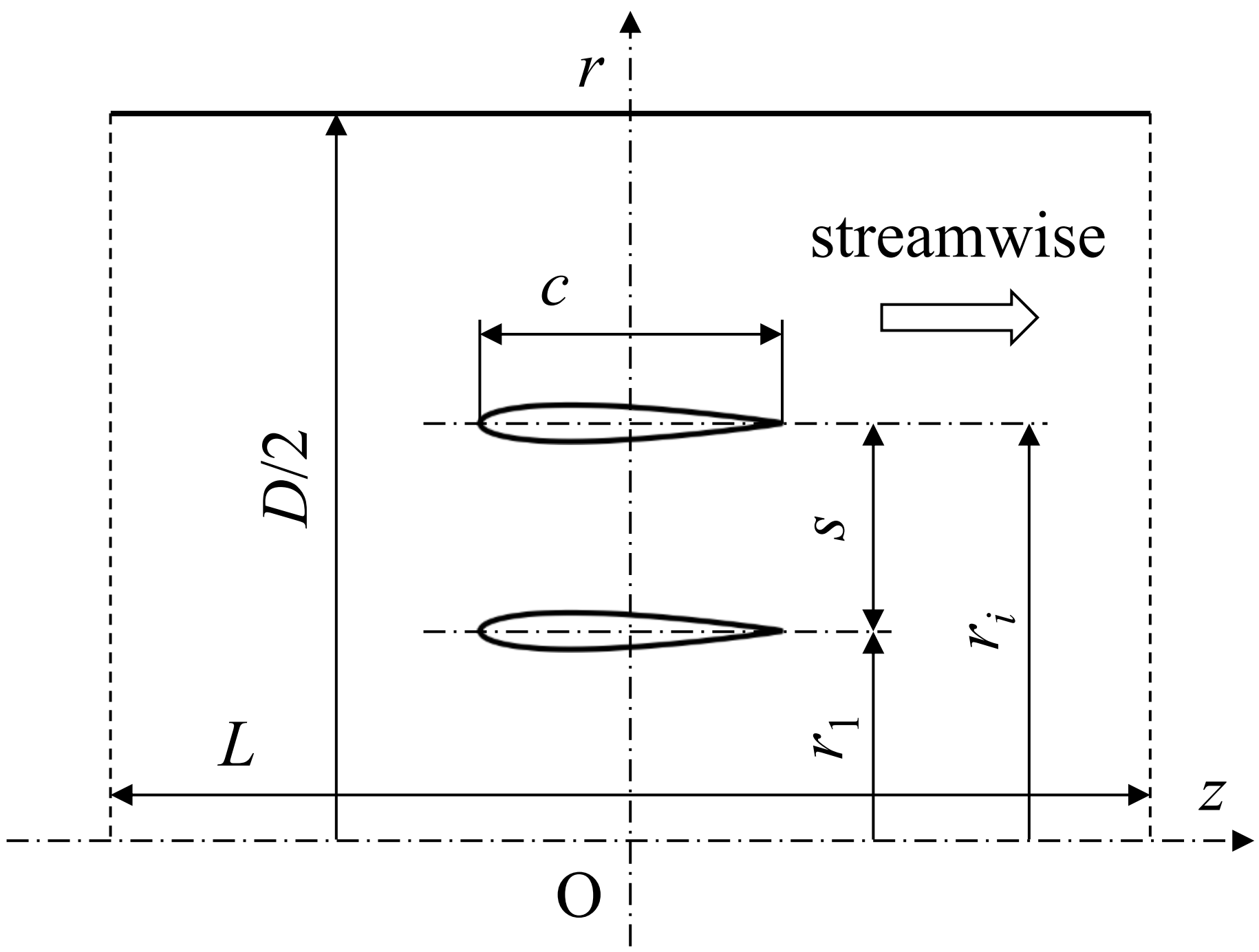

FIG. 1. Schematic diagram of the flow straightener within the turbulent flow pipe in the streamwise direction. Mathematical notation is defined in the text.

The working fluid is an incompressible Newtonian fluid, and steady turbulent flow is assumed through Reynolds averaging. A modified version of the $k$–$\omega$ shear stress transport (SST) model[31,32] was adopted as the RANS model with the following governing equations:

$$\nabla \cdot \mathbf{U} = 0, \tag{2}$$

$$\nabla \cdot (\mathbf{U} \otimes \mathbf{U}) = -\frac{1}{\rho}\nabla P + \nabla \cdot [(\nu + \nu_t)(\nabla \mathbf{U} + (\nabla \mathbf{U})^T)], \tag{3}$$

$$\nabla \cdot (\mathbf{U}k) = P_k - \beta^* k\omega + \nabla \cdot [(\nu + \sigma_k \nu_t)\nabla k], \tag{4}$$

$$\nabla \cdot (\mathbf{U}\omega) = \alpha \frac{\omega}{k} P_k - \beta\omega^2 + \nabla \cdot [(\nu + \sigma_\omega \nu_t)\nabla \omega] + 2(1 - F_1)\sigma_{\omega 2}\frac{1}{\omega}\nabla k \cdot \nabla \omega, \tag{5}$$

$$P_k = \nu_t 2S_{ij}S_{ij}, \qquad S_{ij} = \frac{1}{2}\left(\partial_i U_j + \partial_j U_i\right), \tag{6}$$

$$\nu_t = \frac{a_1 k}{\max(a_1\omega,\ SF_2)}, \tag{7}$$

$$F_1 = \tanh(\arg_1^4), \qquad \arg_1 = \min\left[\max\left(\frac{\sqrt{k}}{\beta^*\omega y}, \frac{500\nu}{y^2\omega}\right), \frac{4\rho\sigma_{\omega 2}k}{CD_{k\omega}y^2}\right], \tag{8}$$

$$F_2 = \tanh(\arg_2^2), \qquad \arg_2 = \max\left(\frac{2\sqrt{k}}{\beta^*\omega y}, \frac{500\nu}{y^2\omega}\right), \tag{9}$$

$$CD_{k\omega} = \max\left(2\rho\sigma_{\omega 2}\frac{1}{\omega}\nabla k \cdot \nabla\omega, 10^{-10}\right), \tag{10}$$

where **U** is the Reynolds-averaged velocity, $P$ is the modified pressure, $\rho$ is the density, $\nu$ is the kinematic viscosity, $\nu_t$ is the eddy viscosity, $S$ is the strain-rate magnitude, and $y$ is the distance to the nearest wall. The model constants were set to $\beta^* = 0.09$, $a_1 = 0.31$, $\beta_1 = 3/40$, $\beta_2 = 0.0828$, $\alpha_1 = 5/9$, $\alpha_2 = 0.44$, $\sigma_{k1} = 0.85$, $\sigma_{k2} = 1$, $\sigma_{\omega 1} = 0.5$, and $\sigma_{\omega 2} = 0.856$. These coefficients were blended using $F_1$:

$$\phi = F_1\phi_1 + (1 - F_1)\phi_2, \tag{11}$$

where $\phi$ denotes a model constant ($\alpha$, $\beta$, $\sigma_k$, or $\sigma_\omega$), and the subscripts 1 and 2 denote the values in the inner and outer regions, respectively.

The boundary conditions were as follows. The pipe walls and grid were no-slip, and the near-wall region was resolved down to the viscous sublayer ($y^+ < 1$) on both, with the eddy viscosity set to zero at the walls (nutLowReWallFunction in OpenFOAM). A periodic boundary condition was imposed in the streamwise ($z$) direction. In addition, axisymmetry was assumed in the azimuthal ($\theta$) direction, which reduced the problem to two dimensions.

The driving force was given by the pressure difference between the two ends of the pipe section. The Darcy–Weisbach equation relates the pressure difference to the mean velocity:

$$\frac{\Delta p}{L} = \frac{\lambda}{D} \cdot \frac{1}{2} \rho U_m^2. \tag{12}$$

Because $k$ is periodic in the streamwise direction, the difference in the modified pressure $P$ over one period equals that in the static pressure $p$. The pipe friction coefficient $\lambda$ (= $4C_f$) for a fully developed turbulent flow in a circular pipe is obtained from the Prandtl–Kármán universal law of pipe friction:

$$\frac{1}{\sqrt{\lambda}} = 2 \log_{10}\left(Re\sqrt{\lambda}\right) - 0.8. \tag{13}$$

When $\lambda$ is known, the wall shear stress is given by

$$\tau_w = \frac{\lambda}{8} \rho U_m^2. \tag{14}$$

If the grid reduces the drag, the flow rate $Q$ should increase even under the same pressure difference. Therefore, the performance of a grid pattern was evaluated by the relative flow rate increase $FR$, which was calculated from the flow rates with and without the grid:

$$FR = Q_{\text{with}}/Q_{\text{without}} - 1. \tag{15}$$

OpenFOAM v2212 was used as the solver. The governing equations were discretized by the finite volume method. A second-order upwind scheme was used for the convection terms, and second-order central differencing was used for the other terms. Pressure and velocity were coupled by the SIMPLE algorithm. The inner iterations were continued until the L1 norm of the residual fell below an absolute tolerance of $10^{-8}$ or a relative tolerance of 0.01. Convergence of the outer iterations was confirmed by monitoring the flow rate until it remained unchanged to four significant digits. The RANS model constants were left at their default values.

As shown in Fig. 2, the mesh around an airfoil was shaped into a slender rectangular block assuming $c$ = 1 mm and $s$ = 1 mm. The streamwise length of the mesh block was 10$c$ upstream and 50$c$ downstream of the airfoil, and its radial length was equal to $s$. This block was fitted into the mesh of the pipe body, which had a different structure, and the two were joined.

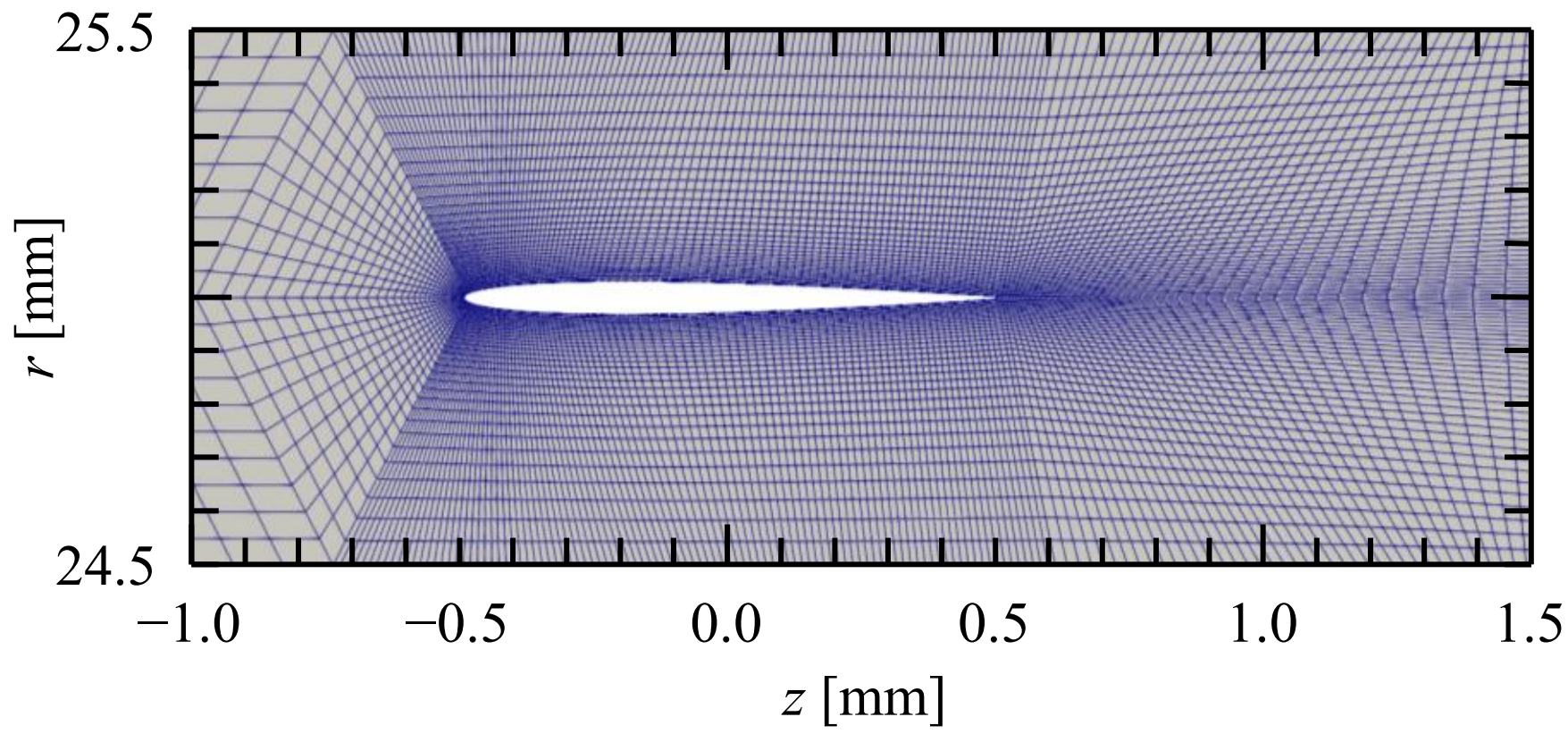

FIG. 2. Mesh resolution around the airfoil.

## III. VERIFICATION AND VALIDATION

### A. Mesh dependence for fully developed pipe flow

To determine the mesh dependence of the RANS model, a fully developed turbulent flow was computed, and the reproduction accuracy was evaluated. Because the flow is uniform in the streamwise direction, the problem can be reduced to one dimension. When the flow rate is fixed so that $Re = 10^5$, Eqs. (13) and (14) give $\lambda = 0.01799$ and $\tau_w/\rho = 0.002249$ m$^2$/s$^2$. The number of cells in the $r$ direction was set to 200, and the cell size was varied geometrically so that the cells were dense near the wall. The effect of the minimum cell height $\Delta r_{\min}$ adjacent to the pipe wall was then investigated by varying the ratio of the maximum to minimum cell sizes. Table I summarizes $\Delta r_{\min}$ [mm], $y^+$, $\tau_w/\rho$ [m$^2$/s$^2$], and the error relative to $\tau_w/\rho$ obtained from Eq. (14). For $y^+ \leq 0.3$, the error was estimated to be 1% or less. When the driving force is the pressure difference, the pressure difference required to achieve $Re = 10^5$ with a pipe length $L = 5$ m is $\Delta p/\rho = 0.4498$ m$^2$/s$^2$ according to Eq. (12). The mean velocity is then expected to be $U_m = 1$ m/s. When the cells in the central part of the pipe were made equally spaced to reduce the number of cells ($\Delta r_{\min}$ =

0.00900 mm), the model gave $U_m$ = 1.003 m/s. Considering that the Prandtl–Kármán universal law of pipe friction given by Eq. (13) itself contains an error, the accuracy was judged to be sufficient. In addition, taking the ratio of flow rates with and without the grid canceled the error originating from the pipe mesh to some extent.

TABLE I. Mesh dependence of the RANS model for a fully developed pipe flow.

| $\Delta r_{\min}$ [mm] | $y^+$ | $\tau_w/\rho$ [m$^2$/s$^2$] | Error |
|---|---|---|---|
| 0.01378 | 0.3246 | 0.002219 | −1.32% |
| 0.01155 | 0.2728 | 0.002232 | −0.76% |
| 0.00948 | 0.2245 | 0.002243 | −0.25% |

B. Reproduction of a previous experiment

The RANS model was used to reproduce the experiment of Mah et al.[25] in which a LEBU device was installed in a circular pipe to confirm that the model can adequately calculate the flow around a LEBU device. Mah et al. installed a cylindrical LEBU device with a length of 46 mm and thickness of 0.3 mm in a circular pipe with a length of more than 10 m and an inner diameter $D$ = 92 mm. Both ends of the cylinder were sharpened. In the simulation, periodic boundary conditions were imposed instead of inflow and outflow boundary conditions, and the pipe length was set to $L$ = 10 m. The minimum mesh size around the LEBU device was set equal to that at the pipe wall. For the comparison, the number of LEBU devices was set to one, and $Re = 6.5 \times 10^4$ was adopted. Three cases were examined where $d$ corresponds to the diameter of the LEBU device: $d/D$ = 0.22, 0.50, and 0.65.

Table II indicates that the difference in $FR + 1$ between the experiment and simulation was 0.2% or less. Figure 3 compares the ratio of local skin friction coefficients $C_f/C_{f0}$ at the pipe wall with and without the LEBU device. While a quantitative discussion is difficult because the experimental error was 0.05 or more, the qualitative features of the experiment and simulation agree. The ratio $C_f/C_{f0}$ decreased locally downstream of the LEBU device, and this tendency was more pronounced at larger $d/D$. These results confirmed that the RANS model reproduced the flow rate within the experimental uncertainty and the local skin friction qualitatively, which indicates that it can also be applied to the flow around the airfoil grid.

TABLE II. Flow rate ratio $FR + 1$ for Mah et al.'s experiment and simulation in the present study.

| $d/D$ | $FR + 1$ (Mah et al.) | $FR + 1$ (Present) | Difference |
|---|---|---|---|
| 0.22 | 99.92% ± 0.2% | 99.75% | −0.17% |
| 0.50 | 99.64% ± 0.2% | 99.60% | −0.04% |
| 0.65 | 99.58% ± 0.2% | 99.65% | 0.07% |

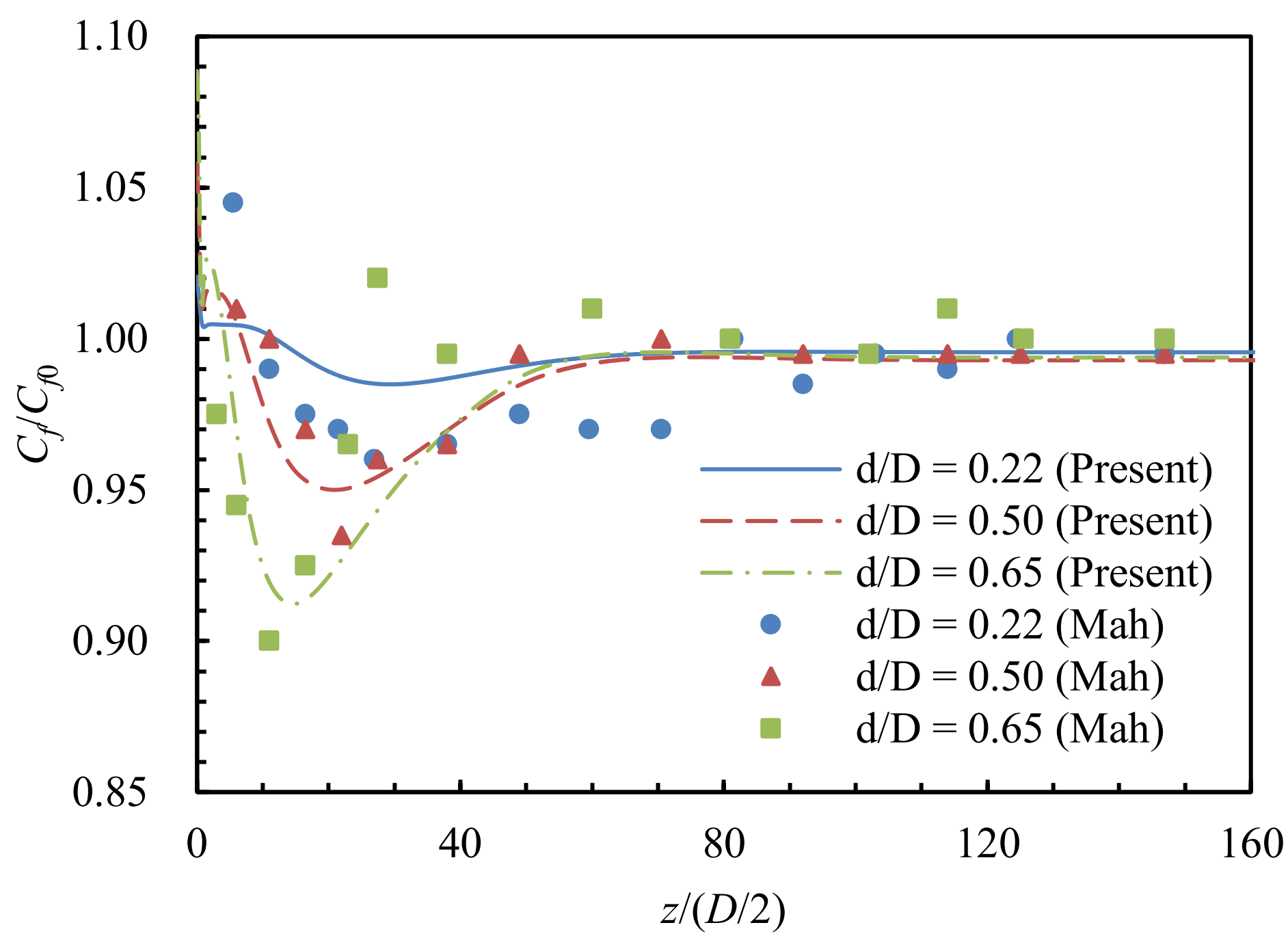

FIG. 3. Experimental (Mah) and simulated (Present) local skin friction coefficient ratios $C_f/C_{f0}$ with and without the LEBU device.

C. Comparison of RANS models

To examine the dependence of the simulation results on the RANS model, the $k$–$\omega$ SST model was compared against other RANS models implemented as standard in OpenFOAM. Five models were tested on the one-dimensional problem specified in Sec. III A: $k$–$\varepsilon$, RNG $k$–$\varepsilon$, realizable $k$–$\varepsilon$, $k$–$\omega$, and $k$–$\omega$ SST. The wall shear stress derived from the Prandtl–Kármán universal law of pipe friction (Eq. (13)) was $\tau_w/\rho$ = 0.002249 $m^2/s^2$. The above models obtained values of 0.004696 (error: 108.78%), 0.004471 (98.80%), 0.000204 (−90.93%), 0.002261 (0.55%), and 0.002243 $m^2/s^2$ (−0.25%), respectively. These calculations were continued for 10,000 iterations. Figure 4 shows the evolution of the integral of the eddy viscosity $\langle \nu_t \rangle = \int_0^{D/2} \nu_t dr$. For the three $k$–$\varepsilon$-type models, $\langle \nu_t \rangle$ reached a minimum in the middle of the iterations and then increased unnaturally by a factor of approximately three or more. Thus, they were excluded from further consideration. When the standard $k$–$\omega$ model was applied to the experiment in Sec. III B, it gave $FR$ + 1 = 98.94% for $d/D$ = 0.50 (experiment: 99.64%), thus overestimating the pressure loss due to the drag of the LEBU device. These results indicate that only the $k$–$\omega$ SST model gave consistently correct results.

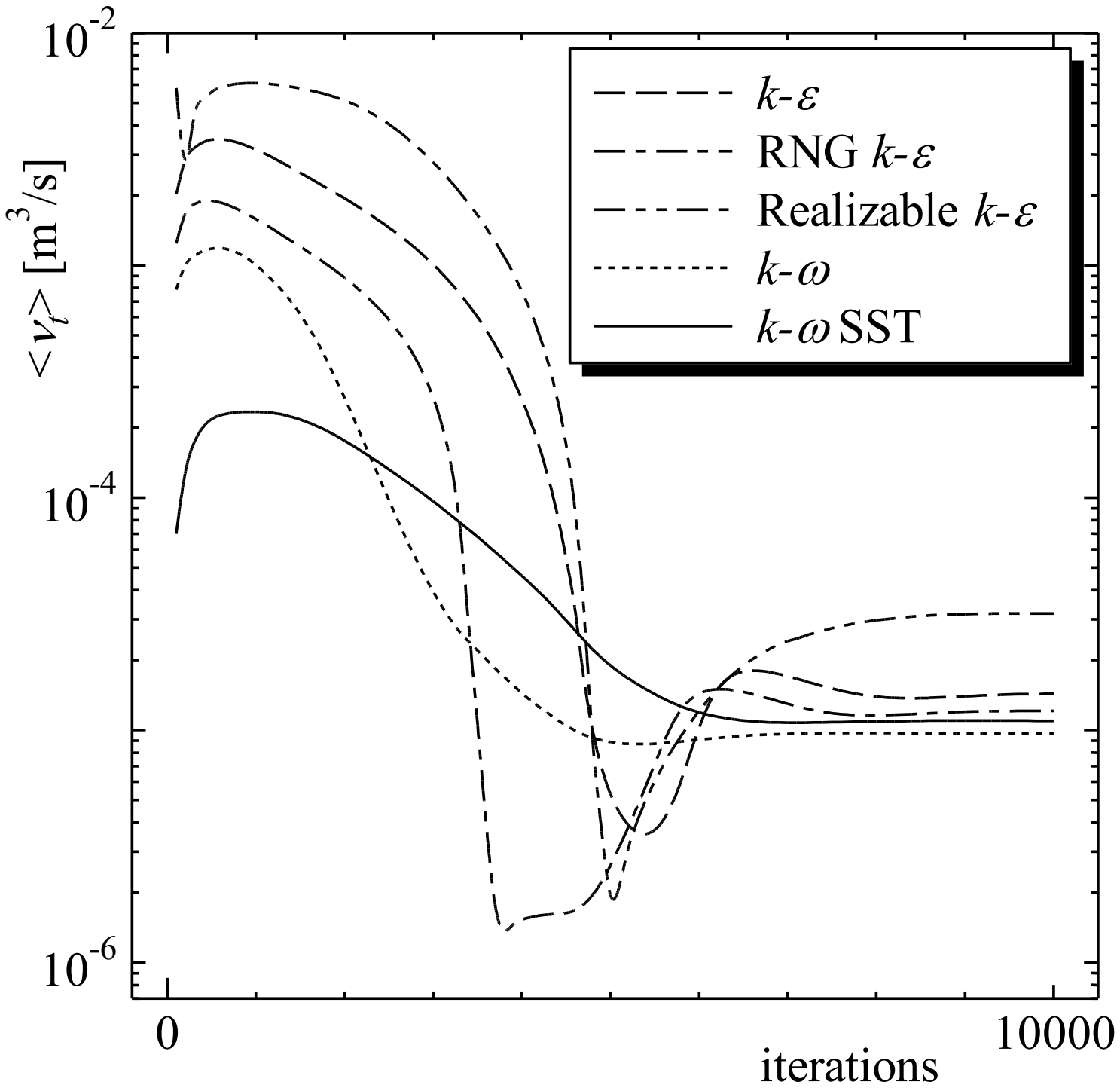

FIG. 4. Evolution of the integral of the eddy viscosity $\langle \nu_t \rangle$ in the one-dimensional problem for different RANS models.

D. Mesh dependence around a single-ring grid

To determine the mesh around the airfoil, it would be preferable to compare the simulation results with experimental results. However, $U_m$ = 1 m/s and $c$ = 1 mm correspond to a chord-based Reynolds number of $Re_c$ = 1,000, which is several orders of magnitude smaller than that of aircraft. Unfortunately, no experimental data could be found for a NACA0006 airfoil installed in a turbulent pipe flow corresponding to $Re_c$ = 1,000. Therefore, the required mesh resolution was evaluated by referring to numerical simulations of airfoils at low Reynolds numbers.[33,34] In particular, Winslow et al.[33] employed a RANS model as in the present study and were a useful reference. Based on their findings, the number of cells along the airfoil was set to 250, and the height of cells adjacent to the airfoil was kept to 0.002$c$ or less. To evaluate the mesh dependence, calculations were performed with the number of cells in the block around the airfoil (Fig. 2) changed to approximately 0.5 and 2 times the baseline.

Table III summarizes the number of cells $N$ in the block, $y^+_{\max}$, the drag per airfoil ring $F_D$ [N], the mean velocity $U_m$ [m/s], and their relative errors. The pipe diameter was $D$ = 100 mm, the kinematic viscosity was $\nu = 10^{-6}$ m$^2$/s, the axial grid spacing (pipe length) was $L$ = 5 m, and the airfoil size was $c$ = 1 mm at $r$ = 25 mm. A pressure difference $\Delta p/\rho = 0.08995L$ [m$^2$/s$^2$] was applied between the two ends of the pipe so that the Reynolds number without the airfoil was $Re = 10^5$. The drag $F_D$ was calculated with a water density of $\rho$ = 1,000 kg/m$^3$. A slight drag reduction was observed with a single-ring grid, and $U_m$ increased slightly from 1 m/s, but the effect of the mesh resolution on the flow rate variation was extremely small. Incidentally, recent RANS simulations of the flow around airfoils have often used transition models,[35,36] but these models are not intended for use with periodic boundary conditions, and the calculation diverges immediately. However, ordinary (nontransition) models estimate the drag of a LEBU device to be higher than transition models do, so any estimations by the model are likely to be conservative.

Figure 5 shows the distribution of the wall shear stress $\tau_w/\rho$ on the pipe wall, which confirms that installing the grid lowered $\tau_w/\rho$ locally. The distribution of $\tau_w/\rho$ changed abruptly at the grid ($z$ = 0 m), but the change was continuous (see inset). The wall shear stress decreased sharply immediately downstream of the grid and reached its minimum at $z$ = 0.86 m. Further downstream, it increased approximately linearly and converged to an approximately constant value for $z \geq 4$ m. This converged value is slightly larger than $\tau_w/\rho$ without the grid because the flow rate increased.

Here, the relationship between the drag and Reynolds stress is briefly discussed. According to the FIK identity (Eq. (1)), suppressing the Reynolds stress is important for reducing drag. As shown in Fig. 6(a), a wake with a somewhat lower Reynolds stress component $R_{rz}/\rho$ formed downstream of the grid. In a RANS model, the Reynolds stress is approximated from the eddy viscosity and velocity gradient using the Boussinesq assumption. To evaluate the wake of a body, the eddy viscosity is often used rather than the Reynolds stress. Figure 6(b) visualizes the eddy viscosity $\nu_t$, where the wake of the grid is clearly visible.

TABLE III. Verification of mesh dependence around the airfoil with the error relative to the next finer mesh.

| $N$ | $y^+_{max}$ | $F_D$ [N] | Error | $U_m$ [m/s] | Error |
|---|---|---|---|---|---|
| 42,014 | 0.6906 | $1.015 \times 10^{-2}$ | −0.011% | 1.010 | 0.000% |
| 82,590 | 0.5068 | $1.015 \times 10^{-2}$ | 0.017% | 1.010 | 0.015% |
| 162,630 | 0.3669 | $1.015 \times 10^{-2}$ | ⋯ | 1.010 | ⋯ |

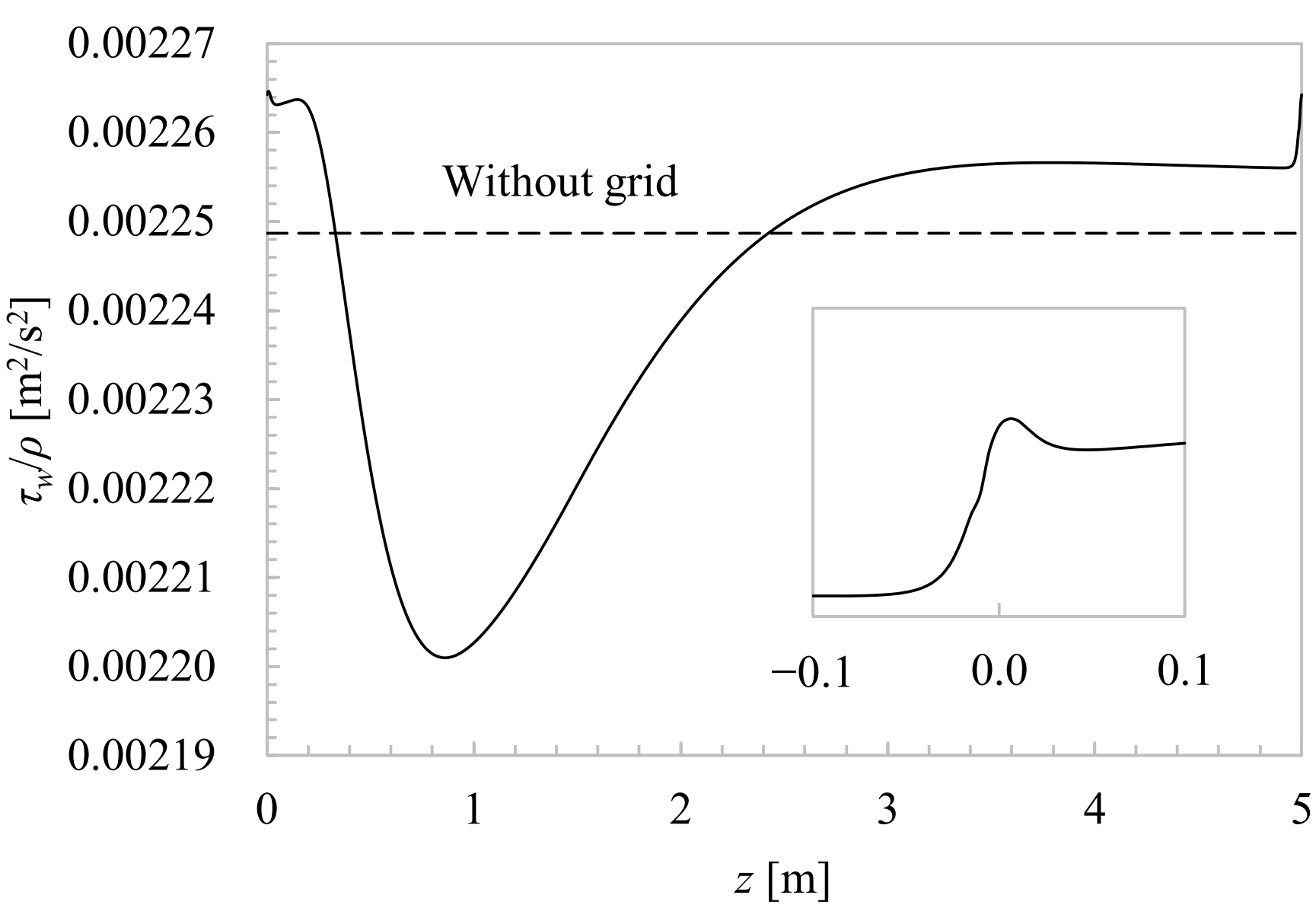

FIG. 5. Distribution of $\tau_w/\rho$ on the pipe wall surface and an enlarged view near $z = 0$ (inset).

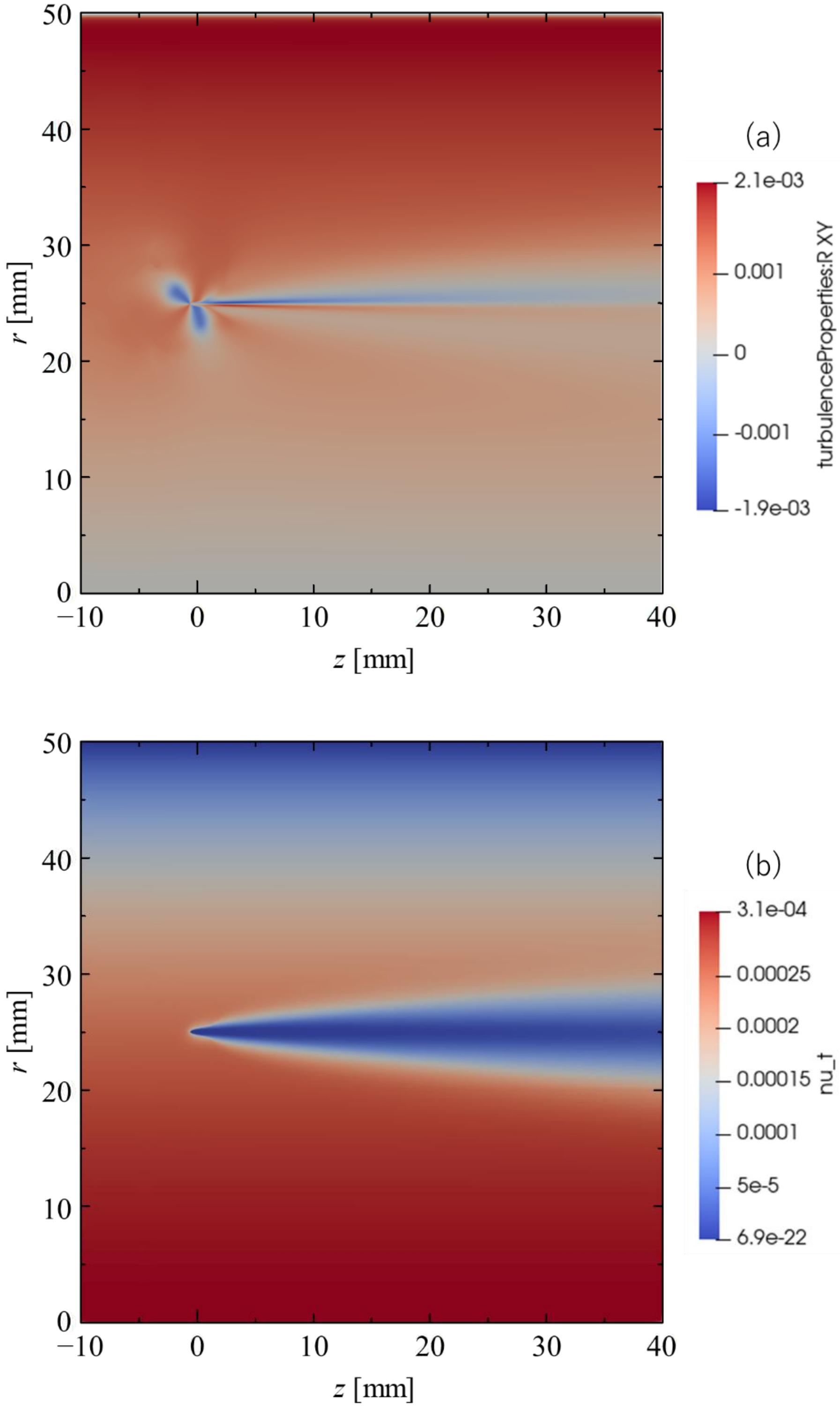

FIG. 6. Distributions of (a) the Reynolds stress component $R_{rz}/\rho$ [$m^2/s^2$] and (b) eddy viscosity $\nu_t$ [$m^2/s$] around a single-ring grid.

# IV. PARAMETER OPTIMIZATION

## A. Chord length

A single-ring grid installed at $r$ = 25 mm was used to investigate the effect of the chord length $c$ on the flow characteristics. The other conditions were the same as those of the baseline case specified in Sec. III D, except that a single-block mesh was used to accommodate larger chord lengths, and *FR* was evaluated relative to the flow without the grid computed with a correspondingly structured mesh. As shown in Fig. 7(a), *FR* varied approximately linearly with $c$, and *FR* tended to be higher for smaller $c$. However, $c$ = 1 mm was adopted because a thinner airfoil would be difficult to fabricate and may not withstand the water pressure. On the other hand, Fig. 7(b) shows that a larger $c$ would be more effective if only the drag from the pipe wall surface was considered. However, the increase in pressure loss (drag) was even larger, and the net drag performance deteriorated. The location where $\tau_w/\rho$ was minimum did not change much with $c$.

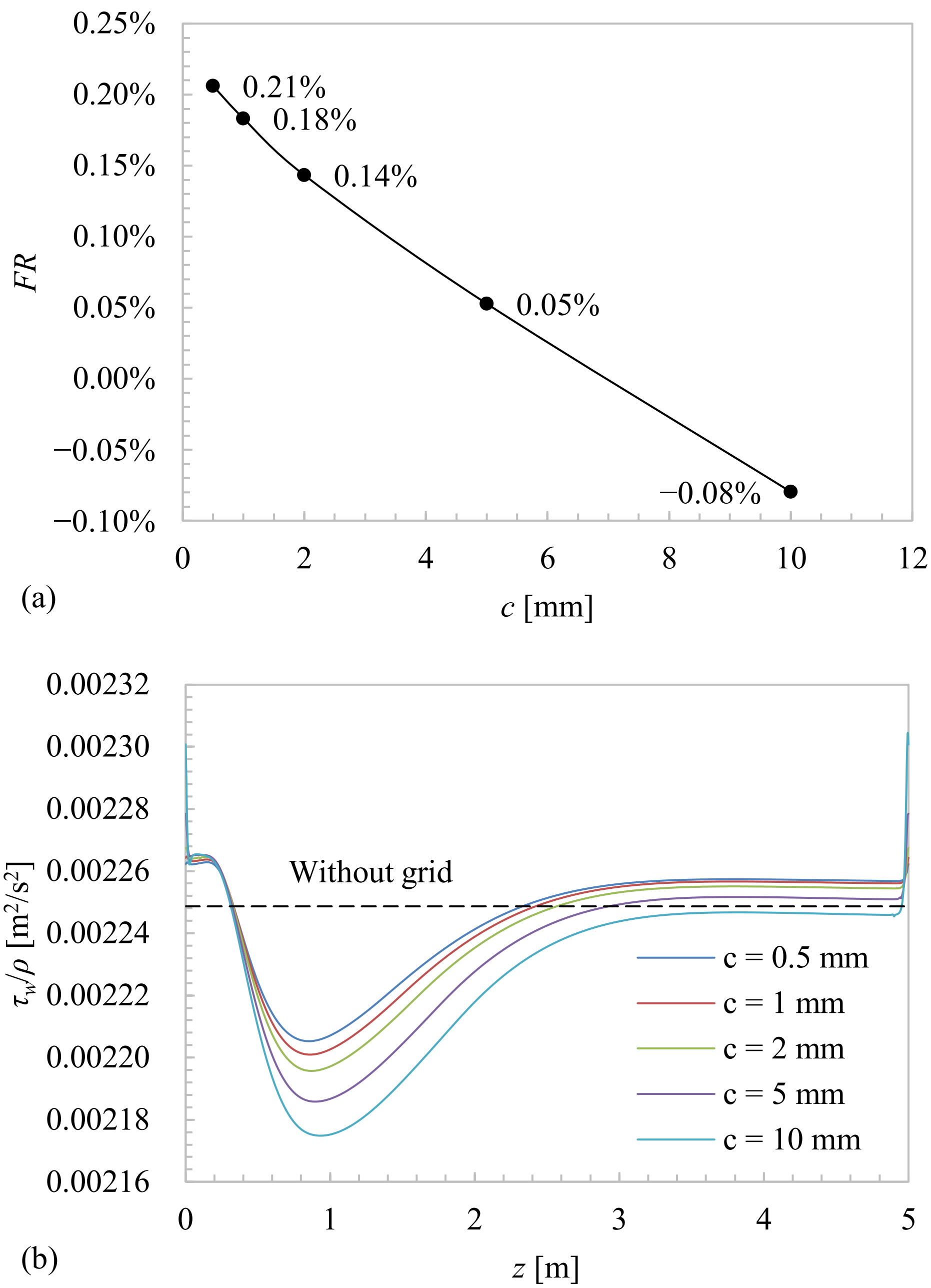

FIG. 7. Distributions of (a) *FR* and (b) $\tau_w/\rho$ with respect to *z* for various *c*. Here, a single-ring grid was placed at $r = 25$ mm.

B. Ring radius

The radius *r* of a single-ring grid was varied while the axial grid spacing was fixed at $L =$ 1 m. As shown in Fig. 8(a), the distribution of *FR* against *r* was a mountain-shaped curve skewed toward the pipe wall, and *FR* was maximum at $r = 38$ mm. This indicates that concentrating the grid near the pipe wall is effective at reducing drag. When the grid was spread over the entire pipe cross-section with $s = 2$ mm, $r_1 = 2$ mm, and $n = 24$, this resulted in $FR = -0.40\%$, which was worse than without the grid. This may be why no net drag reduction by a flow straightener has yet been reported.

The Reynolds shear stress in a fully developed turbulent pipe flow is known to be proportional to the radius except near the pipe wall. Therefore, the second term in Eq. (1) can be regarded as an integral of $r^3$. The FR distribution in Fig. 8(a) can be approximated almost exactly by a cubic polynomial of $r$ in the range $r \leq 30$ mm, which reflects Eq. (1). On the other hand, when an airfoil is located near the pipe wall, it becomes difficult for the fluid to flow between them, which increases the pressure loss.

As shown in Fig. 8(b), the distribution of $\tau_w/\rho$ in the $z$ direction decreased more sharply near the grid with increasing $r$ (i.e., as the grid came closer to the pipe wall) and conversely became flatter with decreasing $r$. Although not shown, the mean value of $\tau_w/\rho$ reached its minimum at $r = 42$ mm. As shown in Fig. 8(c), the $z$ coordinate at which $\tau_w/\rho$ is minimum varied linearly with $r$. If the wake width simply grew in proportion to $z^{1/2}$, the distance at which the wake reaches the pipe wall would be proportional to $(D/2 - r)^2$. The linear dependence therefore suggests that other factors, such as the development of vortices from the pipe wall and the weighting by $r$ in Eq. (1), also play a role.

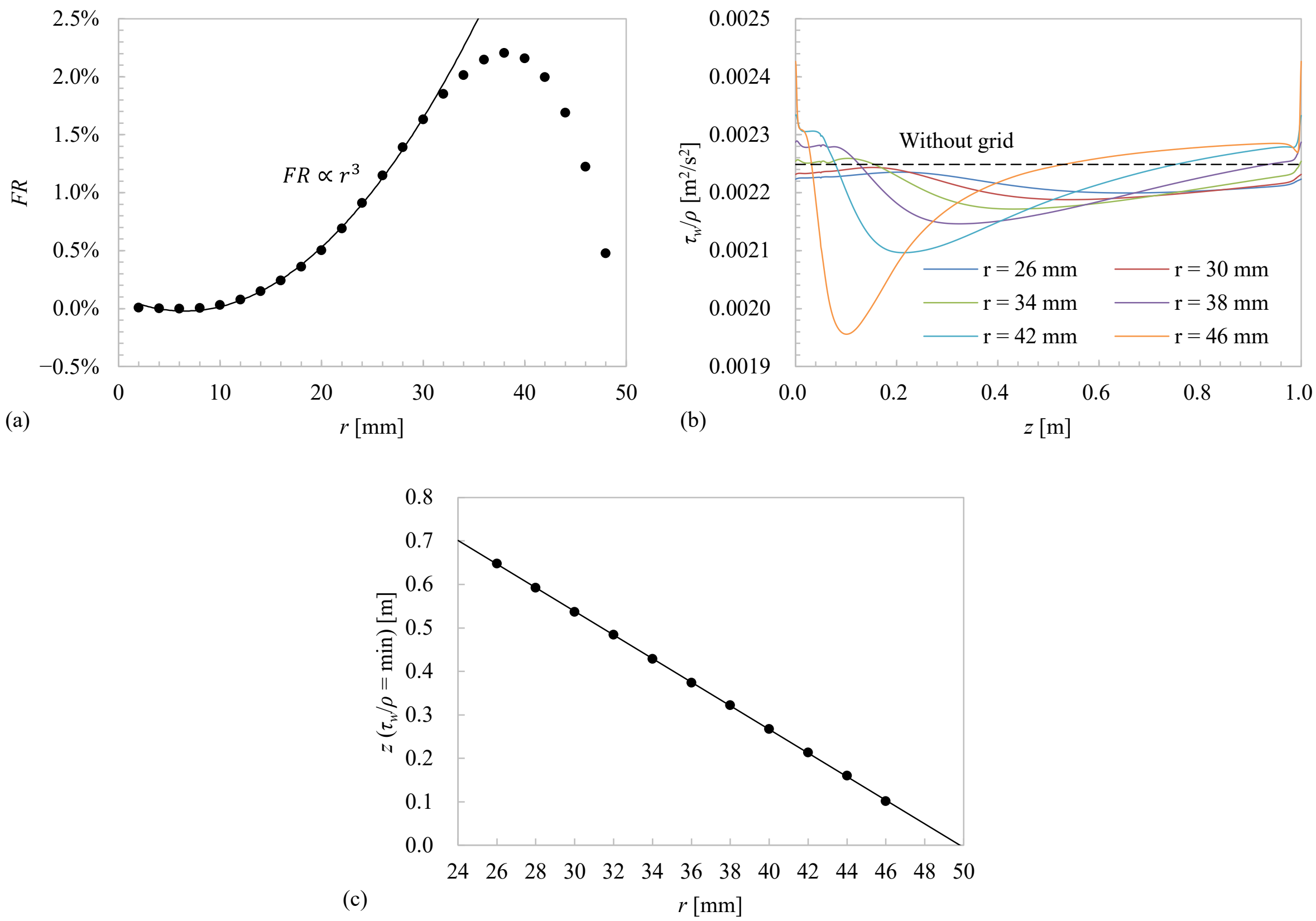

FIG. 8. Distributions of (a) *FR*, (b) $\tau_w/\rho$ with respect to $z$, and (c) the $z$ at which $\tau_w/\rho$ is minimum, for various $r$ of a single-ring grid. Here, $L$ = 1 m.

## C. Radial ring spacing

With the grid concentrated near the pipe wall ($r$ = 25–50 mm), the radial ring spacing $s$ was optimized next. In addition to $s$, the grid pattern is defined by the innermost ring radius $r_1$ and number of rings $n$. Table IV lists the tested grid patterns. *FR* increased the most with $s$ = 3 mm but deteriorated rapidly with smaller $s$, which indicates that care is needed when designing the grid pattern. Figure 9 shows that the $\tau_w/\rho$ distribution was relatively flat in the $z$ direction. It was lowest immediately downstream of the grid and then recovered gradually, which is similar to the sum of the single-ring $\tau_w/\rho$ distributions shown in Fig. 8(b). Here, only $s$ = 5 mm demonstrated slightly different characteristics because this case used a different equation for determining the ring radii.

The following equations were used to calculate the radius of the outermost ring:

$$r_n = D/2 - D/100 - s/2\,, \qquad s = 2, 3, 4, 6. \tag{16}$$

$$r_n = D/2 - s/2\,, \qquad s = 5. \tag{17}$$

The radii of the other rings are given by $r_i = r_n - s(n - i)$.

TABLE IV. Different combinations of $s$, $r_1$, and $n$ tested and the corresponding $FR$.

| $s$ [mm] | $r_1$ [mm] | $n$ | $FR$ |
|---|---|---|---|
| 2 | 26.0 | 12 | 5.66% |
| 3 | 26.5 | 8 | 7.11% |
| 4 | 27.0 | 6 | 6.94% |
| 5 | 27.5 | 5 | 6.35% |
| 6 | 28.0 | 4 | 5.81% |

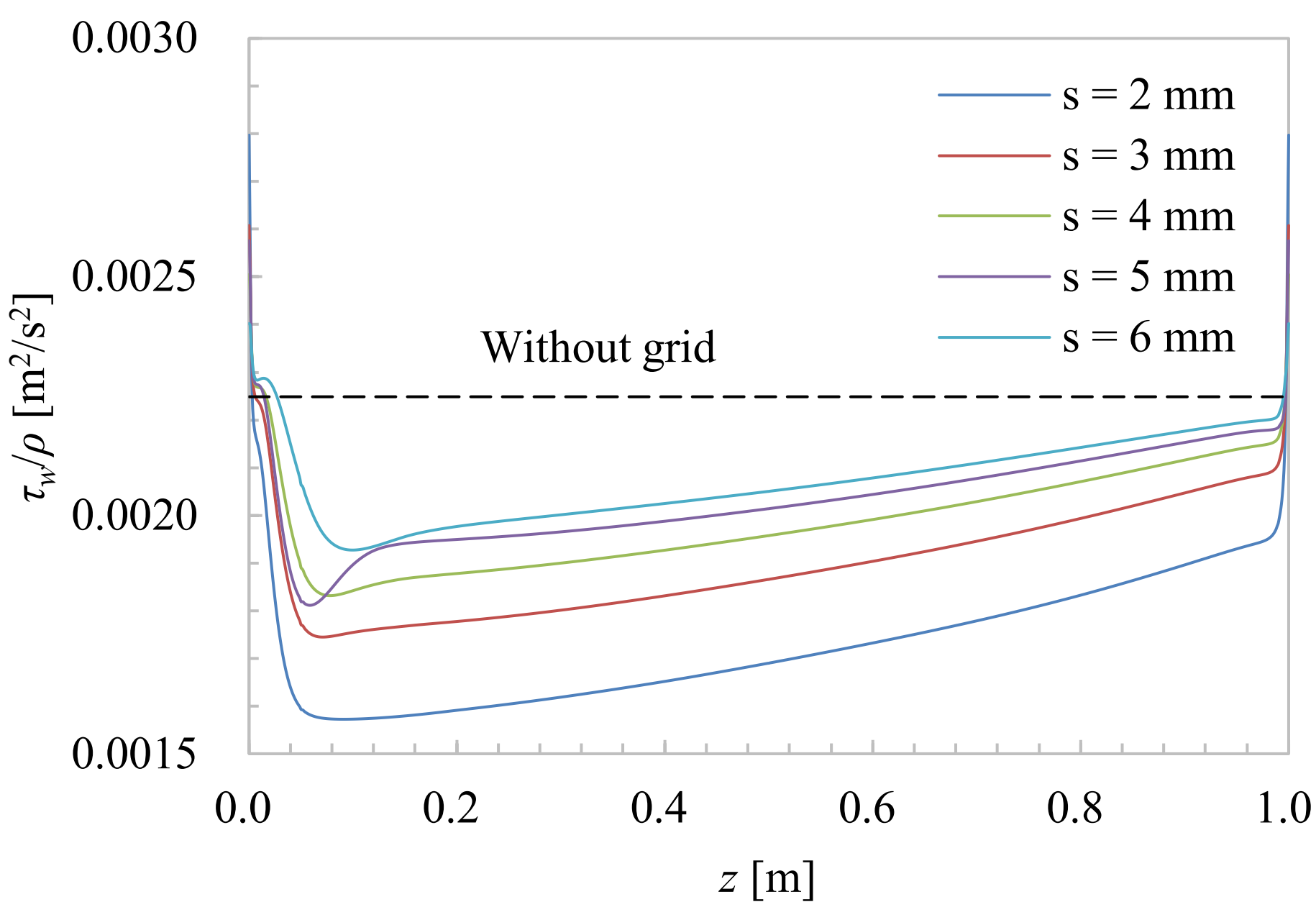

FIG. 9. Distributions of $\tau_w/\rho$ with respect to $z$ for various $s$. Here, $L$ = 1 m, and the grid patterns correspond to those listed in Table IV.

D. Optimal conditions

For a fixed $s$ = 3 mm, the optimal combination of parameters that gave the highest *FR* by adding or removing rings was determined using $r$ = 38 mm as the reference. The results are summarized in Table V. The best combination was $r_1$ = 29.5 mm and $n$ = 7.

This grid pattern was then used to determine the optimal $L$. As shown in Fig. 10(a), the highest *FR* = 9.02% was obtained at $L$ = 0.5 m. Even with the standard $k$–$\omega$ model, which overestimated the drag of the LEBU device in Sec. III C, *FR* for this condition was 3.57%, smaller but still positive. Note that, because periodic boundary conditions were imposed, $L$ represents the spacing of an infinite streamwise array of grids rather than the recovery length downstream of a single grid. Care is required because *FR* deteriorates rapidly if $L$ is too short. Figure 10(b) shows the distribution of $\tau_w/\rho$ along the $z$ axis for the best grid pattern with varying $L$. These profiles were roughly similar in shape.

TABLE V. Combinations of $r_1$ and $n$ for $s$ = 3 mm and the corresponding *FR*.

| $r_1$ [mm] | $n$ | *FR* |
|---|---|---|
| 23.5 | 9 | 6.75% |
| 26.5 | 8 | 7.11% |
| 29.5 | 7 | 7.23% |
| 32.5 | 6 | 6.95% |
| 29.5 | 6 | 6.76% |

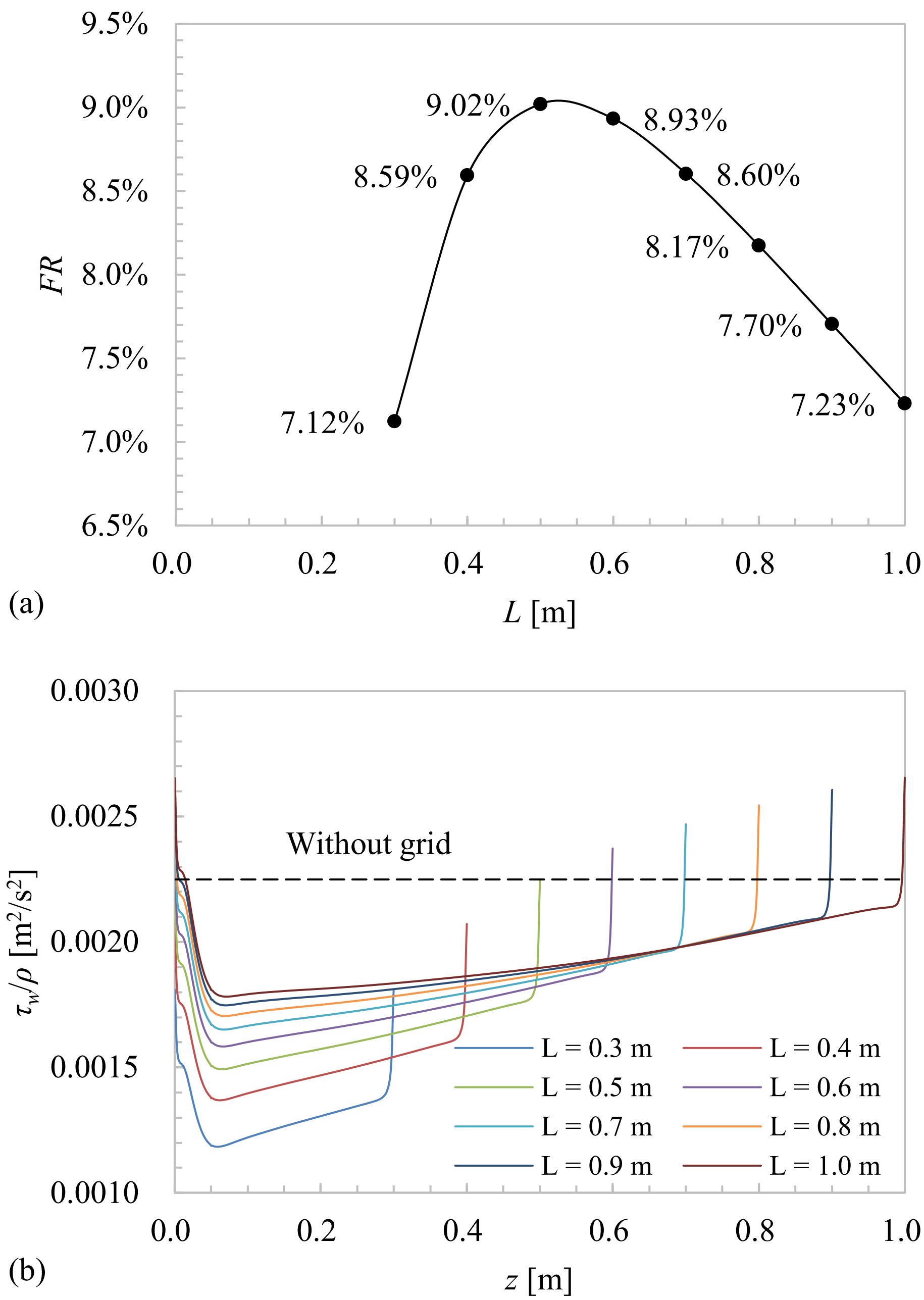

FIG. 10. (a) *FR* and (b) $\tau_w/\rho$ with respect to $z$ for the optimal grid pattern ($s$ = 3 mm, $r_1$ = 29.5 mm, and $n$ = 7) with varying $L$. Here, $D$ = 100 mm, $\nu = 10^{-6}$ m$^2$/s, $c$ = 1 mm, and $\Delta p/\rho = 0.08995L$ [m$^2$/s$^2$] ($Re \approx 10^5$).

E. Flow velocity

The optimal grid pattern found in Sec. IV D was used to investigate the effect of the flow velocity based on changing the pressure difference between the two ends of the pipe. Note that *Re* is the value defined without the grid. As shown in Fig. 11, the $L$ at which *FR* was maximum decreased with increasing *Re*. At $Re = 2 \times 10^5$, *FR* was maximum at $L = 0.4$ m, whereas at $Re = 1 \times 10^4$, the maximum is presumed to occur at $L > 1$ m. For a given $L$, *FR* increased with increasing *Re*. In the range $5 \times 10^4 \leq Re \leq 2 \times 10^5$, *FR* was positive for all $L$. This tendency is favorable for practical use but requires the grid to have higher mechanical strength.

Because the optimal grid pattern is likely to depend on the flow rate, *FR* should increase further if the grid pattern is tailored to a specific *Re*. Nevertheless, Eq. (12) shows that the required $\Delta p/\rho$ is proportional to $U_m^2$, so the pressure loss is already very small at low *Re*. Therefore, it is sufficient to determine the optimal grid pattern and axial grid spacing for the maximum *Re* expected during operation.

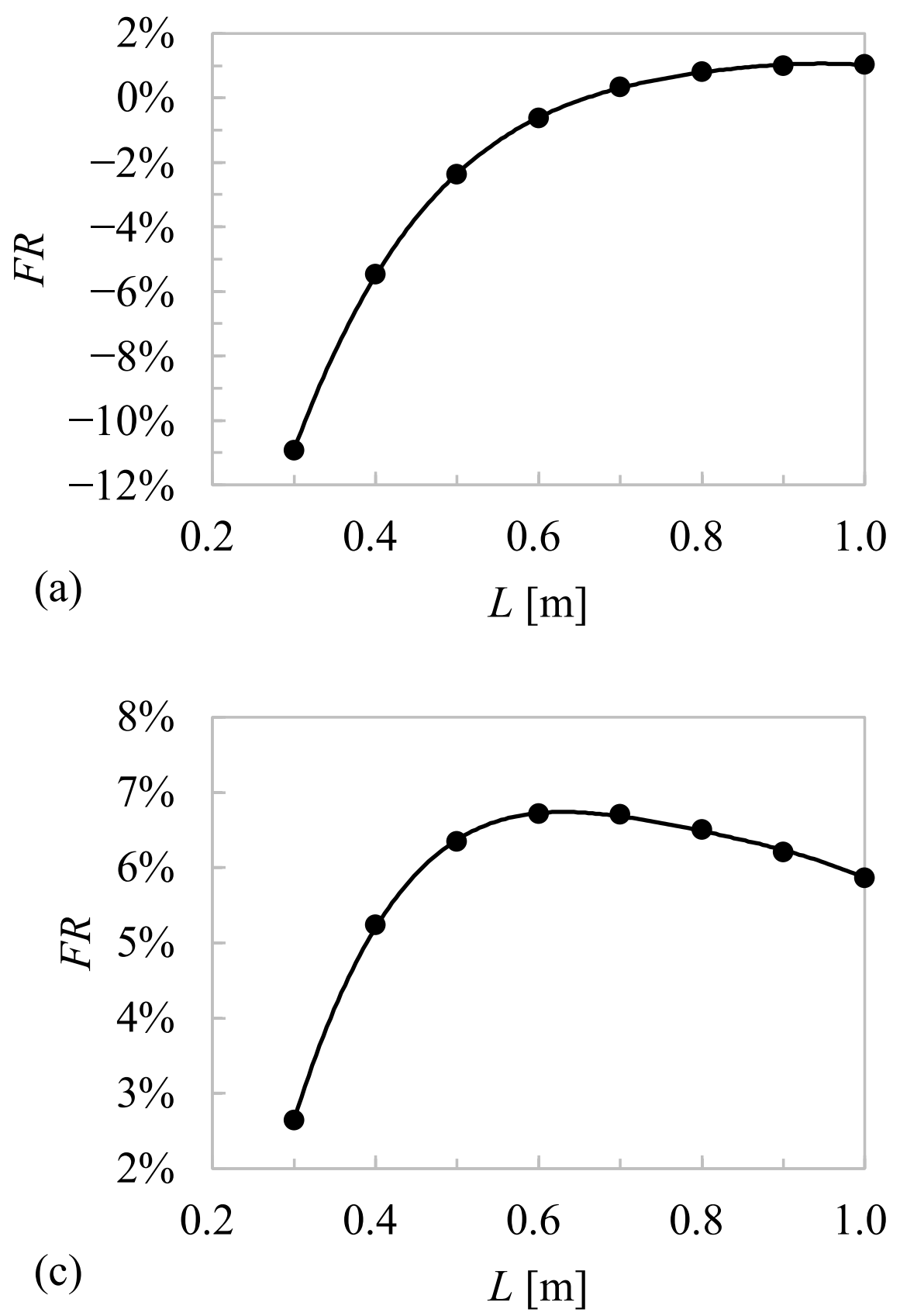

FIG. 11. Distributions of *FR* with respect to *L* at (a) $Re = 1 \times 10^4$, (b) $2 \times 10^4$, (c) $5 \times 10^4$, and (d) $2 \times 10^5$.

# V. MECHANISM

## A. Laminarization

The optimal grid pattern presented in Sec. IV D was used to elucidate the principle of net drag reduction by a flow straightener. As mentioned in Sec. III D and as shown in Fig. 12, the optimal grid pattern results in a wake region with low $R_{rz}/\rho$ and low $\nu_t$ extending downstream around the pipe wall ($r \gtrsim D/4$). The area integral of $R_{rz}/\rho$ is defined as

$$\langle R_{rz}/\rho \rangle_S = \int_{r=0}^{D/2} \int_{z=0}^{L} (R_{rz}/\rho) dr dz. \tag{18}$$

For a given *Re*, $\langle R_{rz}/\rho \rangle_S = 3.21898 \times 10^{-5}\ \mathrm{m^4/s^2}$ without the grid and $\langle R_{rz}/\rho \rangle_S =$ $2.29559 \times 10^{-5}\ \mathrm{m}^4/\mathrm{s}^2$ with the grid, which corresponds to a reduction to 71.3% (by 28.7%).

The distributions of the Reynolds-averaged velocity and pressure were examined. As shown in Fig. 13(a), compared with fully developed turbulent flow at the same flow rate, installing the grid caused the distribution of the velocity component $U_z$ to approach a parabolic profile. This relaxed the velocity gradient $\partial U_z/\partial r$ near the wall, and the wall shear stress $\tau_w = -\mu(\partial U_z/\partial r)|_{r=D/2}$ decreased accordingly. A naive interpretation of this result is that the flow was laminarized and approached the Hagen–Poiseuille solution, which would reduce the flow resistance.

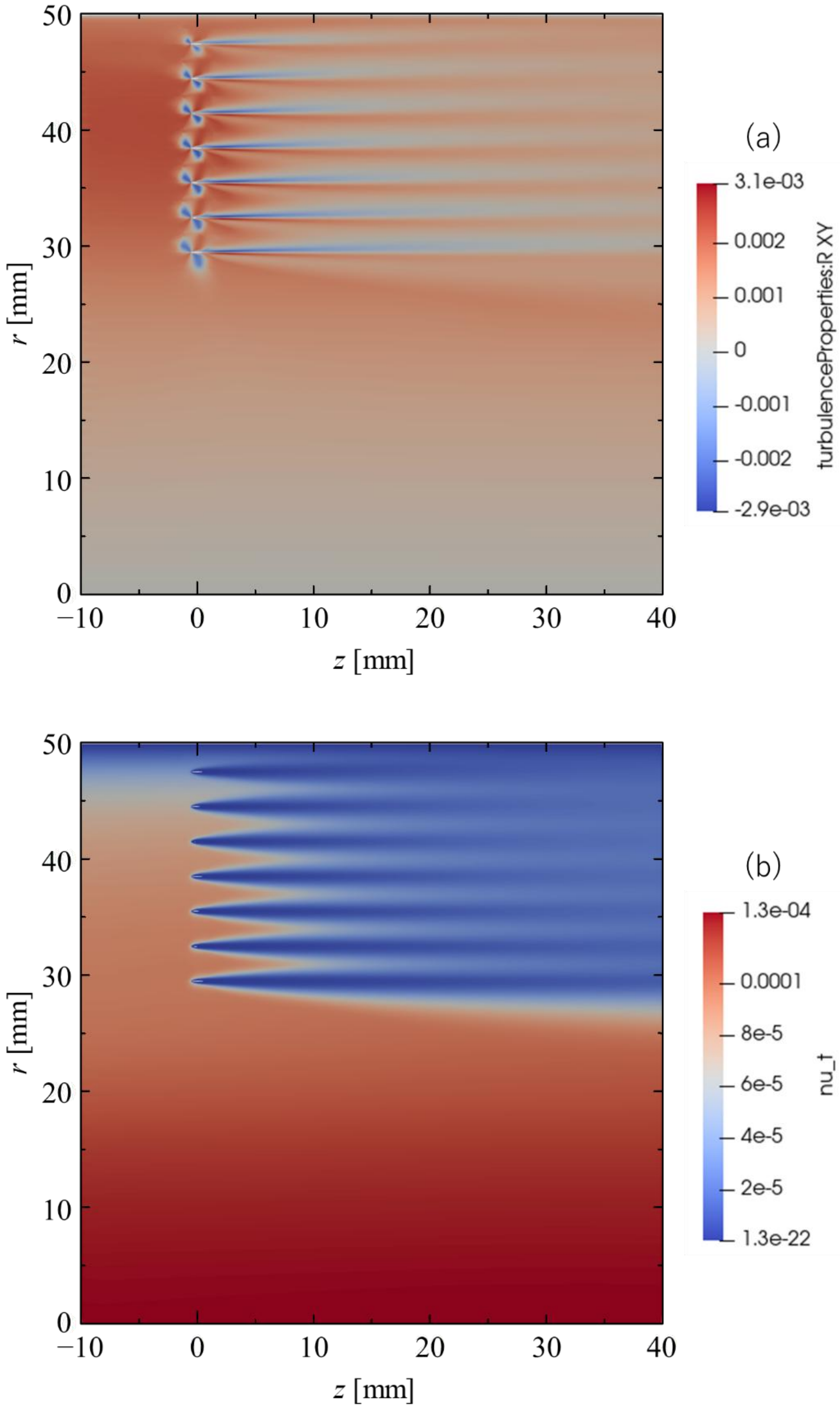

FIG. 12. Distributions of (a) the Reynolds stress component $R_{rz}/\rho$ [m$^2$/s$^2$] and (b) eddy viscosity $\nu_t$ [m$^2$/s] around the grid.

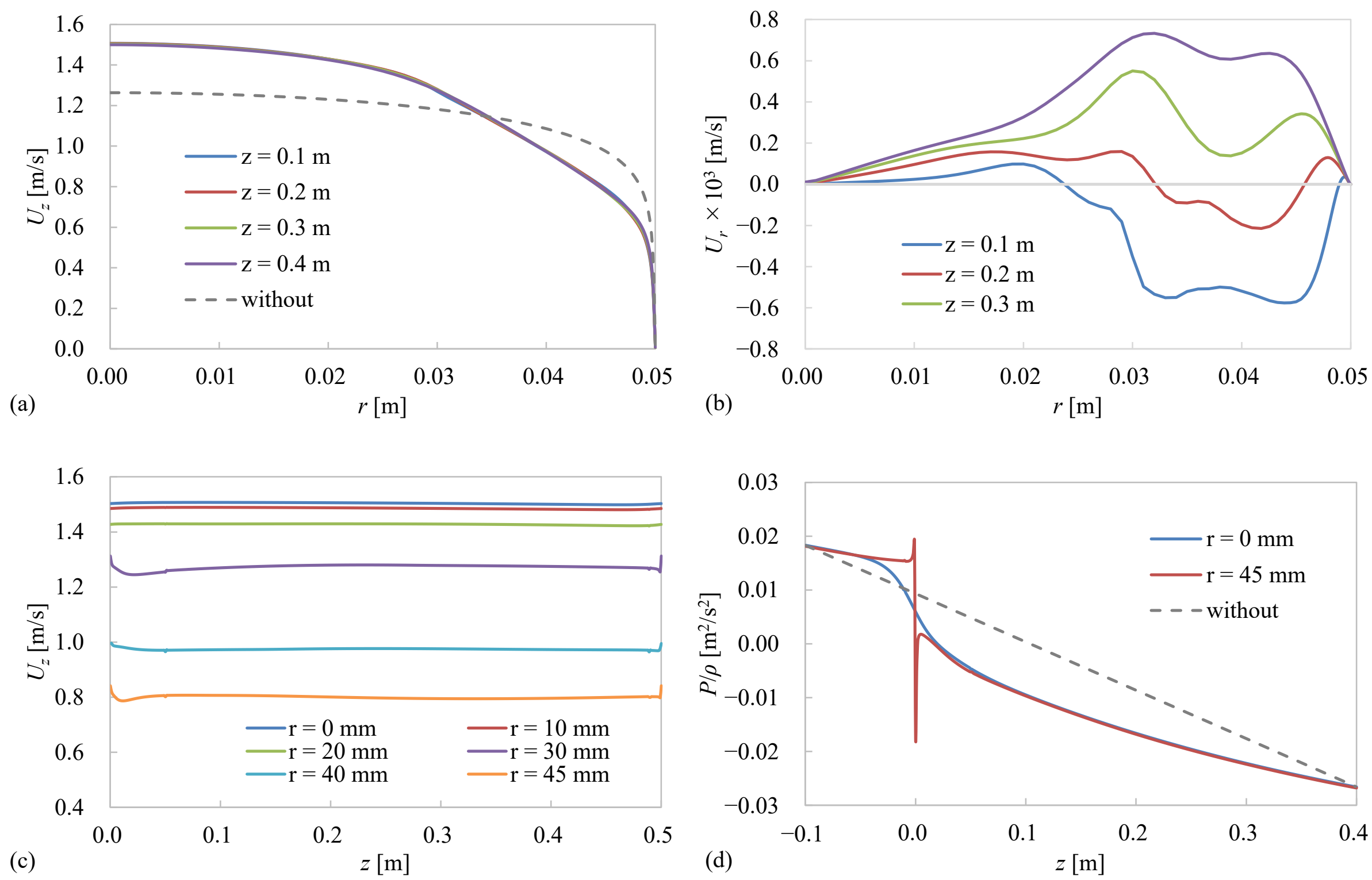

FIG. 13. Distributions of velocity and pressure. (a) $U_z$ in the $r$ direction, (b) $U_r$ in the $r$ direction, (c) $U_z$ in the $z$ direction, and (d) $P/\rho$ in the $z$ direction. In (a), the label "without" denotes a fully developed flow at the same flow rate. In (d), the label "without" denotes a fully developed flow at the same pressure difference.

## B. FIK identity decomposition

Equation (1) gives the FIK identity for a fully developed flow in a circular pipe. For a flow in a circular pipe that is not fully developed, the generalized FIK identity is defined as follows:[30]

$$C_f = \underbrace{\frac{16}{Re_b}}_{\mathrm{DT}} + \underbrace{16\int_0^1 2r^*\overline{u_r^{*\prime}u_z^{*\prime}}r^*dr^*}_{\mathrm{RS}} - \underbrace{16\int_0^1 (r^{*2}-1)\left(I_z'' + \frac{\partial p^{*\prime\prime}}{\partial z^*} + \frac{\partial \overline{u_z^*}}{\partial t^*}\right) r^* dr^*}_{\mathrm{SD}}, \quad (19)$$

$$\overline{I_z} = \frac{1}{r^*}\frac{\partial(r^*\overline{u_r^* u_z^*})}{\partial r^*} + \frac{\partial(\overline{u_z^* u_z^*})}{\partial z^*} - \frac{1}{Re_b}\frac{\partial^2 \overline{u_z^*}}{\partial z^{*2}}. \quad (20)$$

Here,

$$f(r^*,\theta,z^*,t^*) = \bar{f}(r^*,z^*,t^*) + f'(r^*,\theta,z^*,t^*), \tag{21}$$

$$\bar{\bar{f}}(z^*,t^*) = 2\int_0^1 \bar{f}(r^*,z^*,t^*)r^*dr^*, \tag{22}$$

$$f''(r^*,z^*,t^*) = \bar{f}(r^*,z^*,t^*) - \bar{\bar{f}}(z^*,t^*). \tag{23}$$

The nondimensionalization is as described for Eq. (1), except that $U_m$ and $Re_b$ are based on the actual flow rate with the grid. The generalized FIK identity can also be applied to flows in which a LEBU device is installed.[18] In Eq. (19), the first term on the right-hand side corresponds to the laminar term (DT), the second corresponds to the turbulent term (RS), and the third corresponds to the spatial development term (SD), following Chan et al.[18]

Figure 14 shows the local variation in the contribution of each term to $C_f$. The dashed lines represent each term for a fully developed turbulent flow at the same flow rate. As expected, the decrease in the RS term accounted for most of the reduction in $C_f$, but the contribution of the SD term was nonnegligible further downstream of the grid (0.2 m $\lesssim z \leq$ 0.5 m). However, because the SD term was initially positive immediately downstream of the grid, its overall contribution was small.

As shown in Eqs. (19) and (20), the SD term has five components. However, because a steady turbulent flow was assumed in this study, $\partial\overline{u_z^*}/\partial t^* = 0$. In addition, Fig. 13(d) shows that the pressure gradient was approximately linear except near the grid, which contradicts the $z$-dependence of the SD term. This indicates that the contribution of $\partial p^{*\prime\prime}/\partial z^*$ is minor. Therefore, the SD term is dominated by $I_z$. However, the third component of $I_z$ is made negligible by being divided by $Re_b \approx O(10^5)$. In addition, Fig. 13(c) shows that the $U_z$ distribution is flat in the streamwise direction, so the influence of the second component should also be small. In other words, the first term in Eq. (20) accounts for most of the SD term in Eq. (19). This is also supported by Fig. 13(b), in which the $U_r$ profile at $z$ = 0.3 m is roughly an upside-down version of that at $z$ =

0.1 m. If the first term of $I_z$ is related to the flow structure, Fig. 13(b) shows that the flow meanders slightly. Immediately downstream of the grid, a flow from the pipe wall toward the pipe axis is formed. Further downstream, a flow in the opposite direction is formed. However, because $r$ acts in a complicated way in the SD term through its cube, the effect is not determined simply by the $U_r$ distribution alone.

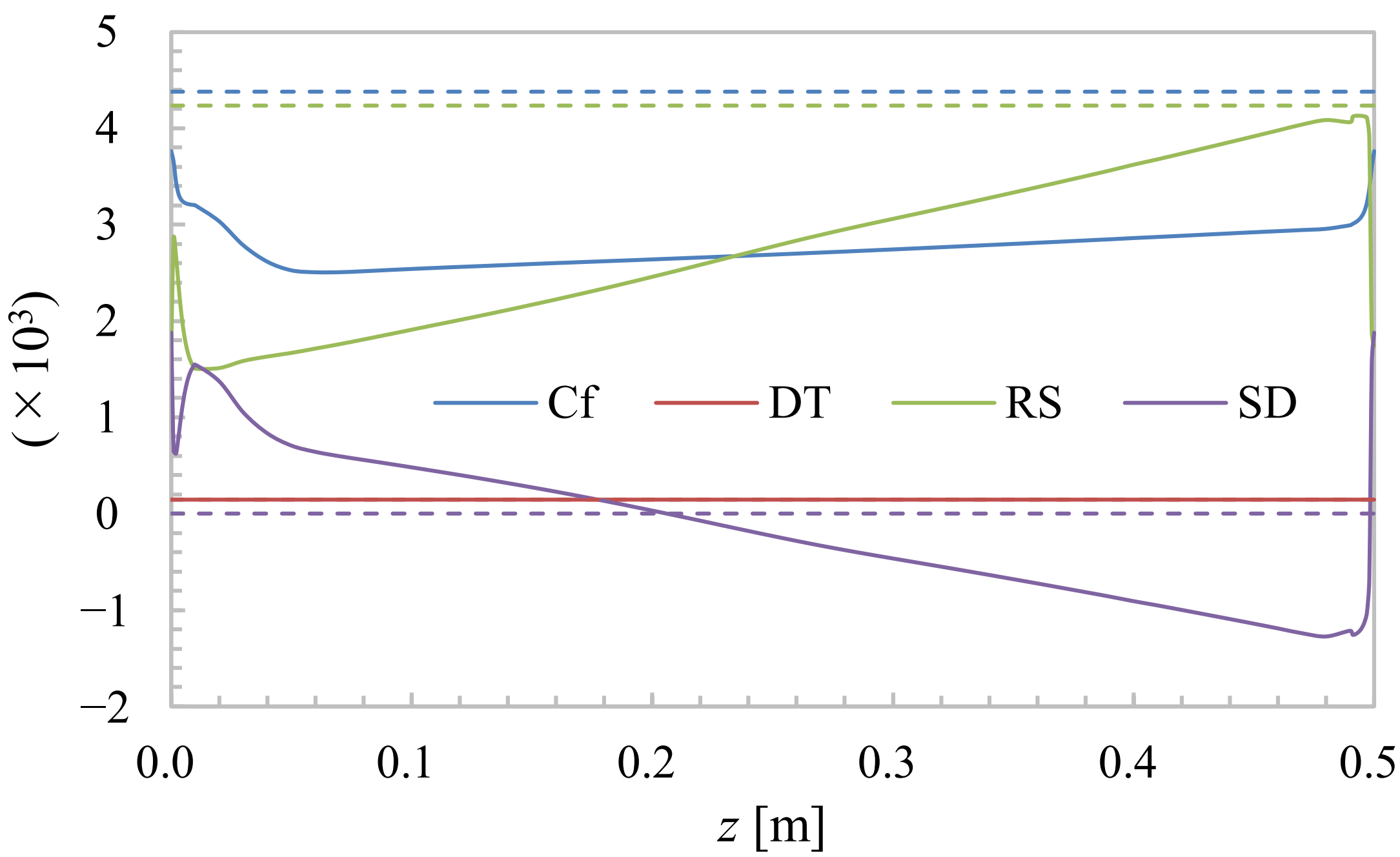

FIG. 14. Contribution of each term of the FIK identity (DT, RS, and SD) to the local skin friction coefficient $C_f$. The dashed lines represent the fully developed flow at the same flow rate.

## VI. CONCLUSIONS

The results showed that placing a flow straightener in a turbulent pipe flow can achieve net drag reduction by using the NACA0006 airfoil as the cross-sectional shape in the streamwise direction to suppress the pressure loss. Under the assumption of a water supply pipe where the driving force is the pressure difference, the net drag reduction of different grid patterns was evaluated in terms of the relative flow rate increase $FR$ at $Re \approx 10^5$. The simulation results showed that the chord length of the airfoil should be minimized. Among single-ring grids, a radius of 38 mm was optimal, and the optimal radial ring spacing was 3 mm. The optimal axial grid spacing depended on the grid pattern but was generally 1 m or less. The optimal grid pattern ($c$ = 1 mm, $s$ = 3 mm, $r_1$ = 29.5 mm, $n$ = 7, and $L$ = 0.5 m) resulted in $FR$ = 9.02% at $Re \approx 10^5$ and retained a positive $FR$ for $5 \times 10^4 \leq Re \leq 2 \times 10^5$. The effectiveness of placing the grid near the pipe wall is consistent with the FIK identity. However, these results were obtained from an axisymmetric RANS model under periodic boundary conditions, and they do not consider the effects of the radial spokes, unsteadiness, or bends and fouling in actual pipes. Verification of the net drag reduction by experiments and DNS/LES, as well as evaluation with three-dimensional geometry, remains for future work. Because drag accounts for a large part of the transport energy in water and gas distribution systems, confirming the effectiveness of the optimal grid pattern at reducing net drag may help these systems reduce their operating costs and environmental impact.

## REFERENCES

1 M. Hino, *Science of Turbulence: Structure and Control* (Asakura Publishing, Tokyo, 2020) (in Japanese).

2 I. Ćipranić, M. Jevrić, M. Radulović, and G. Sekulić, "The energy–pressure nexus in the water supply system," Water Supply **21**(1), 344–355 (2021). https://doi.org/10.2166/ws.2020.285

3 I. Silvestre, R. Pastor, and R. C. Neto, "Power losses in natural gas and hydrogen transmission in the Portuguese high-pressure network," Energy **272**, 127136 (2023). https://doi.org/10.1016/j.energy.2023.127136

4 D. M. Bushnell and C. B. McGinley, "Turbulence control in wall flows," Annu. Rev. Fluid Mech. **21**, 1–20 (1989). https://doi.org/10.1146/annurev.fl.21.010189.000245

5 M. Gad-el-Hak, *Flow Control: Passive, Active, and Reactive Flow Management* (Cambridge University Press, Cambridge, 2000).

6 D. W. Bechert, M. Bruse, W. V. Hage, J. T. Van der Hoeven, and G. Hoppe, "Experiments on drag-reducing surfaces and their optimization with an adjustable geometry," J. Fluid Mech. **338**, 59–87 (1997).
https://doi.org/10.1017/S0022112096004673

7 H. Choi, P. Moin, and J. Kim, "Active turbulence control for drag reduction in wall-bounded flows," J. Fluid Mech. **262**, 75–110 (1994).
https://doi.org/10.1017/S0022112094000431

8 T. R. Bewley, "Flow control: new challenges for a new Renaissance," Prog. Aerosp. Sci. **37**(1), 21–58 (2001).
https://doi.org/10.1016/S0376-0421(00)00016-6

9 P. S. Virk, "Drag reduction fundamentals," AIChE J. **21**(4), 625–656 (1975).
https://doi.org/10.1002/aic.690210402

10 C. M. White and M. G. Mungal, "Mechanics and prediction of turbulent drag reduction with polymer additives," Annu. Rev. Fluid Mech. **40**(1), 235–256 (2008).
https://doi.org/10.1146/annurev.fluid.40.111406.102156

11 M. Quadrio, "Drag reduction in turbulent boundary layers by in-plane wall motion," Philos. Trans. R. Soc. A **369**(1940), 1428–1442 (2011).
https://doi.org/10.1098/rsta.2010.0366

12 P. R. Bandyopadhyay, "Mean flow in turbulent boundary layers disturbed to alter skin friction," J. Fluids Eng. **108**, 127–140 (1986).
https://doi.org/10.1115/1.3242552

13 P. H. Alfredsson and R. Örlü, "Large-eddy breakup devices—a 40 years perspective from a Stockholm horizon," Flow Turbul. Combust. **100**(4), 877–888 (2018).
https://doi.org/10.1007/s10494-018-9908-4

14 P. R. Spalart, M. Strelets, and A. Travin, "Direct numerical simulation of large-eddy-break-up devices in a boundary layer," Int. J. Heat Fluid Flow **27**, 902–910 (2006).
https://doi.org/10.1016/j.ijheatfluidflow.2006.03.014

15 J. S. Kim, J. Hwang, M. Yoon, J. Ahn, and H. J. Sung, "Influence of a large-eddy breakup device on the frictional drag in a turbulent boundary layer," Phys. Fluids **29**, 065103 (2017).
https://doi.org/10.1063/1.4984602

16 C. Chin, R. Örlü, J. Monty, N. Hutchins, A. Ooi, and P. Schlatter, "Simulation of a large-eddy-break-up device (LEBU) in a moderate Reynolds number turbulent boundary layer," Flow Turbul. Combust. **98**, 445–460 (2017).
https://doi.org/10.1007/s10494-016-9757-y

17 C. Chin, R. Örlü, P. Schlatter, J. Monty, and N. Hutchins, "Influence of a large-eddy-breakup-

device on the turbulent interface of boundary layers," Flow Turbul. Combust. **99**, 823–835 (2017).
https://doi.org/10.1007/s10494-017-9861-7

18 I. C. Chan, R. Örlü, P. Schlatter, and R. C. Chin, "The skin-friction coefficient of a turbulent boundary layer modified by a large-eddy break-up device," Phys. Fluids **33**(3), 035153 (2021).
https://doi.org/10.1063/5.0043984

19 C. I. Chan, R. Örlü, P. Schlatter, and R. C. Chin, "Large-scale and small-scale contribution to the skin friction reduction in a modified turbulent boundary layer by a large-eddy break-up device," Phys. Rev. Fluids **7**(3), 034601 (2022).
https://doi.org/10.1103/PhysRevFluids.7.034601

20 Y. Shirose, "Computations of the turbulent frictional drag reduction by LEBU," Trans. West-Jpn. Soc. Nav. Archit. **78**, 21–34 (1989) (in Japanese).
https://doi.org/10.14856/wjsna.78.0_21

21 A. Sahlin, P. H. Alfredsson, and A. V. Johansson, "Direct drag measurements for a flat plate with passive boundary layer manipulators," Phys. Fluids **29**, 696–700 (1986).
https://doi.org/10.1063/1.865921

22 A. Sahlin, A. V. Johansson, and P. H. Alfredsson, "On the possibility of drag reduction by outer layer manipulators in turbulent boundary layers," Phys. Fluids **31**, 2814–2820 (1988).
https://doi.org/10.1063/1.866989

23 T. B. Lynn, D. W. Bechert, and D. A. Gerich, "Direct drag measurements in a turbulent flat-plate boundary layer with turbulence manipulators," Exp. Fluids **19**, 405–416 (1995).
https://doi.org/10.1007/BF00190258

24 A. Pollard, A. M. Savill, and H. Thomann, "Turbulent pipe flow manipulation: some experimental and computational results for single manipulator rings," Appl. Sci. Res. **46**(3), 281–290 (1989).
https://doi.org/10.1007/BF00404825

25 Y. A. Mah, B. C. Khoo, and Y. T. Chew, "The effect of BLADE manipulator in fully developed pipe flow," J. Fluids Eng. **114**(4), 687–689 (1992).
https://doi.org/10.1115/1.2910086

26 B. C. Khoo, Y. T. Chew, and Y. A. Mah, "Skin friction following BLADE manipulation in a turbulent pipe flow," Exp. Fluids **15**(4), 274–278 (1993).
https://doi.org/10.1007/BF00223405

27 G. Yin, M. C. Ong, and P. Zhang, "Numerical investigations of pipe flow downstream a flow conditioner with bundle of tubes," Eng. Appl. Comput. Fluid Mech. **17**(1), e2154850 (2023).
https://doi.org/10.1080/19942060.2022.2154850

28 A. P. Jurga, M. J. Janocha, M. C. Ong, and G. Yin, "Numerical investigations of turbulent flow through a 90-degree pipe bend and honeycomb straightener," J. Fluids Eng. **146**(2), 021307 (2024).
https://doi.org/10.1115/1.4064101

29 J. Kühnen, D. Scarselli, and B. Hof, "Relaminarization of pipe flow by means of 3D-printed shaped honeycombs," J. Fluids Eng. **141**(11), 111105 (2019).
https://doi.org/10.1115/1.4043494

30 K. Fukagata, K. Iwamoto, and N. Kasagi, "Contribution of Reynolds stress distribution to the skin friction in wall-bounded flows," Phys. Fluids **14**(11), L73–L76 (2002).
https://doi.org/10.1063/1.1516779

31 F. R. Menter, "Two-equation eddy-viscosity turbulence models for engineering applications," AIAA J. **32**(8), 1598–1605 (1994).
https://doi.org/10.2514/3.12149

32 F. R. Menter, M. Kuntz, and R. Langtry, "Ten years of industrial experience with the SST turbulence model," Turbul. Heat Mass Transf. **4**(1), 625–632 (2003).

33 J. Winslow, H. Otsuka, B. Govindarajan, and I. Chopra, "Basic understanding of airfoil characteristics at low Reynolds numbers ($10^4$–$10^5$)," J. Aircr. **55**(3), 1050–1061 (2018).
https://doi.org/10.2514/1.C034415

34 D. F. Kurtulus, "Unsteady aerodynamics of a pitching NACA 0012 airfoil at low Reynolds number," Int. J. Micro Air Veh. **11**, 1756829319890609 (2019).
https://doi.org/10.1177/1756829319890609

35 F. R. Menter, R. Langtry, and S. Völker, "Transition modelling for general purpose CFD codes," Flow Turbul. Combust. **77**(1), 277–303 (2006).
https://doi.org/10.1007/s10494-006-9047-1

36 D. K. Walters and D. Cokljat, "A three-equation eddy-viscosity model for Reynolds averaged Navier–Stokes simulations of transitional flow," J. Fluids Eng. **130**(12), 121401 (2008).
https://doi.org/10.1115/1.2979230